\documentclass[aps,prd,12pt,superscriptaddress,nofootinbib]{revtex4-2}
\usepackage{amsmath,amssymb,graphicx,subfigure}
\usepackage{dcolumn}
\usepackage{bm}
\usepackage{soul}
\usepackage{xcolor}

\definecolor{DarkBlue}{rgb}{0.15,0.15,0.85}
\usepackage[linktocpage=true]{hyperref}
\hypersetup{colorlinks=true, citecolor=DarkBlue, linkcolor=magenta, urlcolor=magenta}
\graphicspath{{Figures/}}

\begin{document}
\title{Primordial black holes and secondary gravitational waves from string inspired general no-scale supergravity}
\author{Lina Wu}
\email{wulina@xatu.edu.cn}
\affiliation{School of Sciences, Xi'an Technological University, Xi'an 710021, China}
\author{Yungui Gong}
\email{yggong@hust.edu.cn}
\affiliation{School of Physics, Huazhong University of Science and Technology, Wuhan, Hubei 430074, China}
\author{Tianjun Li}
\email{tli@itp.ac.cn}
\affiliation{School of Sciences, Xi'an Technological University, Xi'an 710021, China}
\affiliation{CAS Key Laboratory of Theoretical Physics, Institute of Theoretical Physics, \\Chinese Academy of Sciences, Beijing 100190, China}
\affiliation{School of Physical Sciences, University of Chinese Academy of Sciences, No.~19A Yuquan Road, Beijing 100049, China}

\begin{abstract}
The formation of primordial black hole (PBH) dark matter and the generation of scalar induced secondary gravitational waves (SIGWs)
have been studied in the generic no-scale supergravity inflationary models. 
By adding an exponential term to the K\"ahler potential, 
the inflaton experiences a period of ultraslow-roll and the amplitude of primordial power spectrum at small scales is enhanced to $\mathcal{O}(10^{-2})$.
The enhanced power spectra of primordial curvature perturbations can have both sharp and broad peaks.
A wide mass range of PBHs can be produced in our model,
and the frequencies of the accompanied SIGWs are ranged form nanohertz to kilohertz. 
We show four benchmark points where the generated PBH masses are around
$\mathcal{O}(10^{-16}M_{\odot})$, $\mathcal{O}(10^{-12}M_{\odot})$, $\mathcal{O}(10^{-2}M_{\odot})$ and $\mathcal{O}(10^{2}M_{\odot})$. 
The PBHs with masses around
$\mathcal{O}(10^{-16}M_{\odot})$ and $ \mathcal{O}(10^{-12}M_{\odot})$ can make up almost all the dark matter, and the accompanied SIGWs
can be probed by the upcoming space-based gravitational wave observatory.
Also, the SIGWs accompanied with the formation of
stellar mass PBHs can be used to interpret the stochastic GW background in the nanohertz band, detected by the North American Nanohertz Observatory for gravitational waves, 
and can be tested
by future interferometric gravitational wave observatory.

\end{abstract}
\maketitle
\section{Introduction}\label{sec:intro}
Since the direct detections of gravitational wave (GW) by the Laser Interferometer Gravitational-Wave Observatory (LIGO) Scientific Collaboration and the Virgo Collaboration \cite{Abbott:2016blz,Abbott:2016nmj,Abbott:2017vtc,Abbott:2017oio,TheLIGOScientific:2017qsa,Abbott:2017gyy,LIGOScientific:2018mvr,Abbott:2020uma,LIGOScientific:2020stg,Abbott:2020khf,Abbott:2020tfl,Abbott:2020niy,LIGOScientific:2021djp}, the idea that primordial black hole (PBH) can be considered as dark matter (DM) candidate \cite{GarciaBellido:1996qt,Leach:2000yw,Bird:2016dcv,Sasaki:2016jop,DeLuca:2020sae,Scholtz:2019csj,Takhistov:2020vxs,DeLuca:2020agl,Vaskonen:2020lbd} has again attracted the attention of physicists and astronomers \cite{Ivanov:1994pa,Frampton:2010sw,Belotsky:2014kca,Khlopov:2004sc,Clesse:2015wea,Carr:2016drx,Inomata:2017okj,Garcia-Bellido:2017fdg,Kovetz:2017rvv,Carr:2020xqk}. The DM fraction in the form of PBHs is tightly constrained by current observations \cite{Carr:2016drx,Carr:2020gox,Green:2020jor}, 
but there is no observational constraint in the mass windows around the masses $10^{-12}M_{\odot}$ and $10^{-16}M_{\odot}$, so
PBHs with masses around $10^{-12}M_{\odot}$ and $10^{-16}M_{\odot}$ can account for the total amount of DM.

PBHs are formed  during radiation domination through gravitational collapse in overdense regions, where the density contrast of small-scale overdense regions at horizon reentry is greater than the threshold value \cite{Hawking:1971ei,Carr:1974nx}. 
To produce PBHs during the radiation era, the primordial scalar power spectrum at small scales should be enhanced to $\mathcal{O}(0.01)$ \cite{Lu:2019sti,Sato-Polito:2019hws}. However, the amplitude of the scalar power spectrum at $k_*=0.05$ Mpc$^{-1}$ is constrained to be $\mathcal{P_{\zeta}}(k_*)=2.10\times10^{-9}$ \cite{Akrami:2018odb,Aghanim:2018eyx}. 
Thus, how to enlarge the primordial curvature perturbation at small scales becomes the key to progress.
One mechanism to realize the enhancement of the power spectrum is the ultraslow-roll inflation with an inflection point \cite{Kawasaki:2016pql,Gong:2017qlj,Garcia-Bellido:2017aan,Garcia-Bellido:2017mdw,Germani:2017bcs,Ezquiaga:2017fvi,Bezrukov:2017dyv,Espinosa:2017sgp,Cicoli:2018asa,Gao:2018pvq,Ballesteros:2018wlw,Dalianis:2018frf,Passaglia:2019ueo,Bhaumik:2019tvl,Xu:2019bdp,Braglia:2020eai}. 
The other mechanisms rely on a peak function either in the nonminimal coupling or in noncannonical kinetic term \cite{Lin:2020goi,Yi:2020kmq,Yi:2020cut,Gao:2020tsa,Gao:2021vxb,Fu:2019ttf,Fu:2019vqc,Dalianis:2019vit,Ballesteros:2020qam,Chen:2020nal,McDonough:2020gmn}. 
The location of the inflection point or the peak in the primordial scalar power spectrum is crucial for the calculation of the mass and abundance of PBHs. 
Near the inflection point, the inflaton potential $V(\phi)$ has an extreme flat plateau where the slow-roll parameter $\epsilon_1$ becomes very small and the power spectrum amplifies. 
However, the slow-roll approximation $|\epsilon_2|\ll 1$ is not satisfied around the point. 
To describe the precise evolution of the power spectrum, the Mukhanov-Sasaki equation should be numerically solved for each mode $k$ \cite{Ballesteros:2017fsr}.
 
Since the scalar perturbations and tensor perturbations are coupled at the nonlinear level, the large primordial curvature perturbations at small scales will induce second-order tensor perturbations after the horizon reentry. The scalar induced secondary GWs (SIGWs) accompanied with the formation of PBH have been extensively studied  \cite{Matarrese:1997ay,Mollerach:2003nq,Ananda:2006af,Baumann:2007zm,Saito:2008jc,Saito:2009jt,Bugaev:2009zh,Bugaev:2010bb,Alabidi:2012ex,Orlofsky:2016vbd,Inomata:2016rbd,Cheng:2018yyr,Cai:2018dig,Bartolo:2018rku,Bartolo:2018evs,Kohri:2018awv,Espinosa:2018eve,Cai:2019amo,Cai:2019bmk,Cai:2019elf,Domenech:2019quo,Cai:2020fnq,Domenech:2020kqm}. 
The large curvature perturbations are the sources of both SIGWs and PBHs, 
and hence GW observations will place limit on the abundance of PBHs. 
The stochastic GW background detected by pulsar timing arrays (PTA) from North American Nanohertz Observatory (NANOGrav) \cite{Arzoumanian:2020vkk} can be explained by SIGWs accompanied with the formation of solar mass PBHs \cite{Vaskonen:2020lbd,Kohri:2020qqd,Nakama:2016gzw}. 
The SIGWs accompanied by PBHs with masses around $\mathcal{O}(10^{-12}M_{\odot})$ can be testable with the space-based GW observatory like Laser Interferometer Space Antenna (LISA) \cite{Danzmann:1997hm}, Taiji \cite{Hu:2017mde}, and TianQin  \cite{Luo:2015ght}.

Inflationary models and PBHs \cite{Ellis:2013nxa, Ellis:2013xoa, Croon:2013ana, Pi:2017gih,Nanopoulos:2020nnh,Ellis:2019bmm,Ellis:2020lnc,Stamou:2021qdk}
have been studied before in the simple no-scale supergravity (one modulus model)~\cite{Cremmer:1983bf, Lahanas:1986uc},
which can be realized via
the Calabi-Yau compactification with standard embedding of
 the weakly coupled heterotic $E_8\times E_8$ theory~\cite{Witten:1985xc} and M-theory on $S^1/Z_2$~\cite{Li:1997sk}.
Moreover, one of us (TL) has studied various orbifold compactifications of M-theory on
$T^6/Z_3$, $T^6/Z_6$, $T^6/Z_{12}$, as well as the compactification
by keeping singlets under $SU(2)\times U(1)$
symmetry, and then the compactification on $S^1/Z_2$~\cite{Li:1998sq}, which provide the general frameworks for
no-scale inflation.
In this paper,  with the general no-scale supergravity theories inspired by the above $T^6/Z_{12}$  orbifold compactification
(three moduli model) and compactification by keeping singlets under $SU(2)\times U(1)$ symmetry (two moduli model)~\cite{Li:1998sq},
we propose the generic inflationary models in which the PBH and SIGWs can be generated. We first discuss the simple and
general no-scale supergravity theories, and show that the inflaton potential in three moduli models is similar to 
the global supersymmetry with canonical K\"ahler potential in Sec. \ref{sec:sugra}, 
but in fact they are completely different due to 
the K\"ahler potential differences.
Then we study the corresponding inflationary models in Sec. \ref{sec:infl}. 
In particular, we find that the tensor-to-scalar ratios in the two and three moduli models~\cite{Li:1998sq} 
are much smaller than that in the simple no-scale model or one modulus model~\cite{Witten:1985xc, Li:1997sk}, 
and thus, the two and three moduli models might provide better frameworks to satisfy the swampland conjecture criteria~\cite{Ooguri:2006in, Obied:2018sgi}. 
The detailed studies will be given elsewhere.
In Sec. \ref{sec:pbh},
we add an exponential term to the K\"ahler potential and show the enhancement of the scalar power spectrum by numerically solving
the Mukhanov-Sasaki equation. 
Then we discuss the production of PBHs and SIGWs.
The conclusions are drawn in Sec. \ref{sec:conclusion}.  
In the following, we set the reduced Planck mass $M_{\text{Pl}}^2=8\pi G=1$.

\section{The Simple and General No-scale supergravity Theories} \label{sec:sugra}
The $\mathcal{N}=1$ supergravity Lagrangian can be written in the form
\begin{equation}\label{eq:lag}
    \mathcal{L}=-\frac{1}{2}R+K_i^{\overline{j}} \partial_{\mu}\varphi^i\partial^{\mu}\overline{\varphi}_{\overline{j}}-V,
\end{equation}
where the K\"ahler metric is $K_i^{\overline{j}}\equiv \partial^2K/(\partial\varphi^i\partial\overline{\varphi}_{\overline{j}})$.  The effective scalar potential is
\begin{equation}
    V=e^G\left[\frac{\partial G}{\partial\varphi^i}\left(K^{-1}\right)^i_{\overline{j}}\frac{\partial G}{\partial \overline{\varphi}_{\overline{j}}}-3\right],
\end{equation}
where the K\"ahler function is $G\equiv K+\ln{|W|^2}$, and $\left(K^{-1}\right)^i_{\overline{j}}$ is the inverse of the K\"ahler metric. 
Introducing the K\"ahler covariant derivative
\begin{equation}
D_i W\equiv  W_i+K_i W,
\end{equation}
the scalar potential can be rewritten as
\begin{equation}
V=e^K\left[D_iW\left(K^{-1}\right)^i_{\overline{j}}D^{\overline{j}}\overline{W}-3|W|^2\right]~.~\,\label{eq:pot-sugra}
\end{equation}


\subsection{The simple no-scale supergravity: One modulus model}
The K\"ahler potential is
\begin{equation}
K=-3 \ln{\left(T_1+\overline{T}_1-|\varphi|^2\right)}.
\end{equation}
The K\"ahler metric and the inverse of the K\"ahler metric are
\begin{equation}
\begin{split}
    K_i^{\overline{j}}=\frac{3}{X^2}\left(
    \begin{array}{cc}
        1 ~&~-\varphi \\
        -\overline{\varphi} ~&~X_1
    \end{array}\right),\\
    \left(K^{-1}\right)^i_{\overline{j}}=\frac{X}{3}\left(
    \begin{array}{cc}
        X_1 ~&~\overline{\varphi} \\
       \varphi ~&~1
    \end{array}\right)~,
\end{split}
\end{equation}
with $X=T_1+\overline{T}_1-|\varphi|^2$ and $X_i= T_i+\overline{T}_i~(i=1,2,3)$.

We consider the superpotential with a  single chiral superfield as
\begin{equation}
\label{eq:superpot}
W=\frac{M}{2}\varphi^2-\frac{\lambda}{3}\varphi^3, 
\end{equation}
which reduces to the Wess-Zumino model with the potential $V=A\phi^2(v-\phi)^2$ when the imaginary part of the scalar component of $\varphi$ vanishes \cite{Croon:2013ana}. In this case, $W_{T_1}=0$ {\footnote{The $\eta$-problem is avoided since no large mass term is generated \cite{Gaillard:1995az,Diamandis:1986zg}.}} and
\begin{equation}
D_i W=W \begin{pmatrix} 
\displaystyle
    -\frac{3}{X} &
\displaystyle    
    \frac{3}{X}\overline{\varphi}+\frac{W_{\varphi}}{W}
\end{pmatrix}.
\end{equation}
Then, the scalar potential becomes
\begin{equation}
\label{eq:pot-Wphi}
V=\frac{|W_{\varphi}|^2}{3X^2}.
\end{equation}

\subsection{The $SU(2)\times U(1)$ symmetry: Two moduli model}

We  consider the compactification of M-theory
by keeping singlets under $SU(2)\times U(1)$
symmetry, and then the compactification on $S^1/Z_2$~\cite{Li:1998sq}, which has two moduli.
The K\"ahler potential is
\begin{equation}
K=-2\log({T_1+\overline{T}_1}-|\varphi|^2)-\log({T_2+\overline{T}_2})~,~
\end{equation}
where for simplicity we neglect the irrelevant scalar fields.
The K\"ahler metric is
\begin{equation}
K_i^{\overline{j}}= \frac{2}{X^2 X_2^2}
    \begin{pmatrix}
        X_2^2 ~&~0 ~&~ - X_2^2 \varphi\\
        0 ~&~X^2/2 ~&~0\\
        - X_2^2\overline{\varphi} ~&~0 ~&~ X_1X_2^2 
    \end{pmatrix}.
\end{equation}
Using the superpotential \eqref{eq:superpot}, the covariant derivative becomes
\begin{equation}
D_i W=W\begin{pmatrix}
\displaystyle -\frac{2 }{X} & \displaystyle-\frac{1}{X_2} & \displaystyle\frac{2}{X} \overline{\varphi}+\frac{W_{\varphi}}{W} \end{pmatrix},
\end{equation}
with $W_{T_i}=0$.
Then the scalar potential becomes
\begin{equation}
V=\frac{|W_{\varphi}|^2}{2X X_2}.
\end{equation}


\subsection{The $T^6/Z_{12}$ orbifold compactification: Three moduli model}

In this subsection, we consider  the $T^6/Z_{12}$ orbifold compactification of M-theory,
 and then the compactification on $S^1/Z_2$~\cite{Li:1998sq}, which has three moduli.
The K\"ahler potential is
\begin{equation}
K=-\log({T_1+\overline{T}_1}-|\varphi|^2)-\sum_{m=2,3}\log({T_m+\overline{T}_m})~,~\,
\end{equation}
where for simplicity we neglect the irrelevant scalar fields as well.
The K\"ahler metric is
\begin{equation}
K_i^{\bar{j}}=\frac{1}{X^2 X_2^2 X_3^2}
\begin{pmatrix}
 X_2^2 X_3^2 & 0 & 0 & -X_2^2 X_3^2\varphi  \\
 0 & X^2 X_3^2 & 0 & 0 \\
 0 & 0 & X^2 X_2^2 & 0 \\
 -X_2^2 X_3^2\overline{\varphi} & 0 & 0 & X_1 X_2^2 X_3^2 \\   
\end{pmatrix}.
\end{equation}
Using the superpotential \eqref{eq:superpot}, the covariant derivative becomes
\begin{equation}
\displaystyle
D_iW=W\begin{pmatrix}
 \displaystyle -\frac{1}{X}&\displaystyle -\frac{1}{X_2}&\displaystyle  -\frac{1}{X_3}& \displaystyle\frac{1}{X}\overline{\varphi}+\frac{W_{\varphi}}{W}
\end{pmatrix},
\end{equation}
with $W_{T_i}=0$.
Then the scalar potential becomes
\begin{equation}
V=\frac{|W_{\varphi}|^2}{X_2 X_3}.
\end{equation}
Thus, after we fix $T_2$ and $T_3$, we obtain the inflationary model similar to that with the global supersymmetry.

\subsection{General parametrization}
To study inflation, PBHs, and SIGWs, we parametrize the generic K\"ahler potential as follows
\begin{equation}
\label{eq:genK}
\begin{split}
K=&-N_X\log({T_1+\overline{T}_1}-|\varphi|^2)-N_Y\log({T_2+\overline{T}_2})\\
&-N_Z\log({T_3+\overline{T}_3}),
\end{split}
\end{equation}
where $N_X+N_Y+N_Z=3$.
In particular, the discussions for inflation, PBHs and SIGWs for the $T^6/Z_{12}$ orbifold compactification are similar to the scenario with $N_X=2$, $N_Y=1$, and $N_Z=0$.
Using the superpotential \eqref{eq:superpot}, the covariant derivative becomes
\begin{equation}
D_iW=W\begin{pmatrix}
\displaystyle -\frac{N_X}{X}&\displaystyle-\frac{N_Y}{X_2} &\displaystyle-\frac{N_Z}{X_3} &\displaystyle \frac{N_X\overline{\varphi}}{X}+\frac{W_{\varphi}}{W}
\end{pmatrix}.
\end{equation}
 Thus, the general scalar potential can be written as
\begin{equation}
V=\frac{|W_{\varphi}|^2}{N_X X^{N_X-1} X_2^{N_Y}X_3^{N_Z}}.
\end{equation}

\section{The Inflationary Models}\label{sec:infl}
We assume that all the real components of the complex fields which do not drive inflation have been stabilized, 
whereas the inflaton field remains dynamical. 
Here we fix the modulus $T_i$ with the vacuum expectation value (VEV) $2\langle {\rm{Re}} (T_i)\rangle = c_i$ and $\langle{\rm{Im}}(T_i)\rangle=0$ \cite{Ellis:2020lnc}, and choose the inflationary trajectory along with $\overline{\varphi}=\varphi$.
For the simple no-scale supergravity, the scalar potential in Eq. \eqref{eq:pot-Wphi} becomes
\begin{equation}
V_1= \frac{M^2\varphi^2(1-d \varphi)^2}{3(c_1-\varphi^2)^2},
\end{equation}
where $d=\lambda/M$. 
One notes that the kinetic term in Eq. \eqref{eq:lag} is noncanonical, 
so we need to define a new canonical field $\chi$, which satisfies
\begin{equation}\label{eq:fieldtrans}
 \frac{1}{2}\partial_{\mu}\chi \partial^{\mu}\chi=K_{\varphi\varphi}\partial_{\mu}\varphi\partial^{\mu}\varphi.
\end{equation}
By integrating the above equation, we get the field transformation $\varphi=\sqrt{c_1}\tanh(\chi/\sqrt{6})$. The Starobinsky model is realized by choosing the specific parameter $ \lambda=\sqrt{3}M$ and $c_1=1/\sqrt{3}$ with the potential
\begin{equation}
\label{Eq:V_oneM}
 V_1= \frac{M^2}{4}\left(1-e^{-\sqrt{\frac{2}{3}}\chi}\right)^2.
\end{equation}

Similarly, the scalar potentials in the two and three moduli models are 
\begin{align}
 V_2= \frac{M^2\varphi^2(1-d \varphi)^2}{2c_2(c_1-\varphi^2)}, \label{eq:v_two}\\
 V_3= \frac{M^2 \varphi ^2 (1- d \varphi )^2}{c_2 c_3}.  
\end{align}
Using the field transformation $\varphi=\sqrt{c_1}\tanh(\chi/\sqrt{2N_X})$, they can be rewritten as
\begin{align}
V_2=&\frac{M^2e^{-\chi} (1-e^{\chi})^2 (1 + e^{\chi} +
	\sqrt{c_1} d (1-e^{\chi}))^2 }{8 c_2 (1 + e^{\chi})^2},\label{Eq:V_twoM}\\
V_3=& \frac{c_1 M^2 \left(1-e^{\sqrt{2} \chi }\right)^2 \left(1+e^{\sqrt{2} \chi }+\sqrt{c_1} d \left(1-e^{\sqrt{2} \chi }\right)\right)^2}{c_2c_3 \left(1+e^{\sqrt{2} \chi }\right)^4}.\label{Eq:V_threeM}
\end{align}

\begin{figure}[htp]
	\centering
	{\includegraphics[height=0.35\columnwidth]{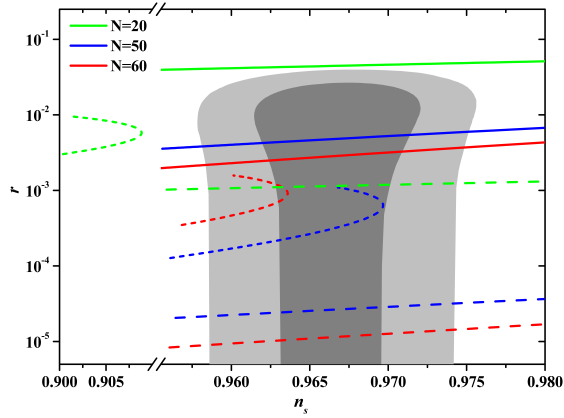}}
	\caption{Observables $n_s$ and $r$ along with constraints from Planck 2018 \cite{Akrami:2018odb,BICEP:2021xfz}. The solid, dashed, and dotted line corresponds to the model with one Eq. \eqref{Eq:V_oneM}, two Eq. \eqref{Eq:V_twoM}, and three Eq. \eqref{Eq:V_threeM} moduli, respectively.}
	\label{fig:nsr}
\end{figure}

The Hubble slow-roll parameters are defined as
\begin{equation}
\epsilon_1=-\frac{\dot{H}}{H^2},~\epsilon_{i+1}=\frac{\dot{\epsilon_i}}{H\epsilon_i},~ i=1,2,3,
\end{equation}
where a dot denotes the derivative with respect to time $t$.
The spectral index $n_s$ and the tensor-to-scalar ratio $r$ in terms of slow-roll parameters are
\begin{equation}
n_s=1+2\epsilon_2-4\epsilon_1,~~r=16\epsilon_1.
\end{equation}
Under the slow-roll condition, we calculate the observables $n_s$ and $r$ with $N=[50,60]$ and the results are shown in Fig. \ref{fig:nsr}. 
The tensor-to-scalar ratio $r$ in these models is much smaller than the Planck 2018 constraint $r_{0.002}<0.056$ (95\% CL) \cite{Akrami:2018odb} and the BICEP/Keck constraint $r_{0.05}<0.036$ (95\% CL) \cite{BICEP:2021xfz}.
Thus, the tensor-to-scalar ratio in the models added by an exponential term in the K\"ahler potential will be tested by the coming GW observatories.
In particular, we find that the tensor-to-scalar ratios in the two and three moduli models \cite{Li:1998sq} 
are much smaller than that in the simple no-scale model or one modulus model \cite{Witten:1985xc, Li:1997sk}, 
so the two and three moduli models might provide better frameworks to satisfy 
the swampland conjecture criteria \cite{Ooguri:2006in, Obied:2018sgi}. The detailed study will be given elsewhere. 

In the next section, we will discuss the enhancement of the primordial power spectrum at small scales by introducing an inflection point in the inflaton potential.
Around the inflection point,
the potential has an extremely flat plateau,
the slow-roll parameter $\epsilon_1$ becomes very small, and the slow-roll parameter $|\epsilon_2|$ becomes large,
so the friction force becomes a driving force and the primordial power spectrum is enhanced,
at the same time the number of $e$-folds also increases dramatically.
To solve the problems of standard big bang cosmology,
inflation has to last $50-60$ $e$-folds.
For the convenience of discussion,
we divide the total number of $e$-folds into two parts: the slow-roll regime $\delta N$ and the ultraslow-roll regime $\Delta N$. The width of the enhanced power spectrum is related to the value of $\Delta N$ or $\delta N$.
If we want to get an enhanced  power spectrum with a broad power spectrum,
the ultraslow-roll regime $\Delta N$ should be bigger and the slow-roll regime $\delta N$ should be less than 30 $e$-folds.
Less slow-roll inflation will make it harder for the model satisfying the constraints from the cosmic microwave background (CMB) observations.  
In Fig. \ref{fig:nsr}, we show the results for $n_s$ and $r$ with $\delta N=20$.
As seen from Fig. \ref{fig:nsr}, either the scalar spectral $n_s$ or the tensor-to-scalar ratio $r$ 
for the models with one and three moduli model is inconsistent with the observational constraints if $\delta N=20$. 
Since the ultraslow-roll $\Delta N$ could have a wide range in the model with two moduli which makes the model easily to realize an enhanced power spectrum without violating the CMB constraints,
thus we discuss PBHs and SIGWs produced in the model with two moduli only.

\section{PBH and SIGWs from the modified K\"ahler potential}\label{sec:pbh}
In this section, we show the enhancement of primordial scalar power spectrum at small scales, 
by adding an exponential term to the K\"ahler potential. 
In this way, an inflection point will be brought into the scalar potential. 
Inflaton will go through a period of ultraslow-roll inflation and the amplitude of the primordial power spectrum is enhanced to $\mathcal{O}(10^{-2})$.  The K\"ahler potential in Eq. \eqref{eq:genK} is modified as
\begin{equation}
\begin{split}
K=&-2\log\left[{T_1+\overline{T}_1}-|\varphi|^2+a e^{-b(\varphi^{\alpha}+\overline{\varphi}^{\alpha})}\left(\varphi^{\beta}+\overline{\varphi}^{\beta}\right)\right]\\
&-\log[{T_2+\overline{T}_2}],
\end{split}
\end{equation}
where $a,~b$ and $\alpha,~\beta$ are real numbers. For simplification, we set  $\alpha=2,~ \beta=2$.  
The parameters $(\alpha,~\beta)$ can also take other integer numbers, 
like $(1,~2)$, and $(2,~2)$. 
The corresponding equations are similar to the model with $\alpha=2,~ \beta=2$, 
so we do not display them.  
The added exponential term differs from that in Refs. \cite{Nanopoulos:2020nnh,Stamou:2021qdk} since the parameter $a$ is a negative real number of $\mathcal{O}(0.1)$ and $b$ is a positive real number of $\mathcal{O}(10)$.  
The correction will not change the whole theory except introducing an inflection point in the effective potential.    

\begin{figure*}
	\centering
	\subfigure[]{\includegraphics[height=0.25\linewidth]{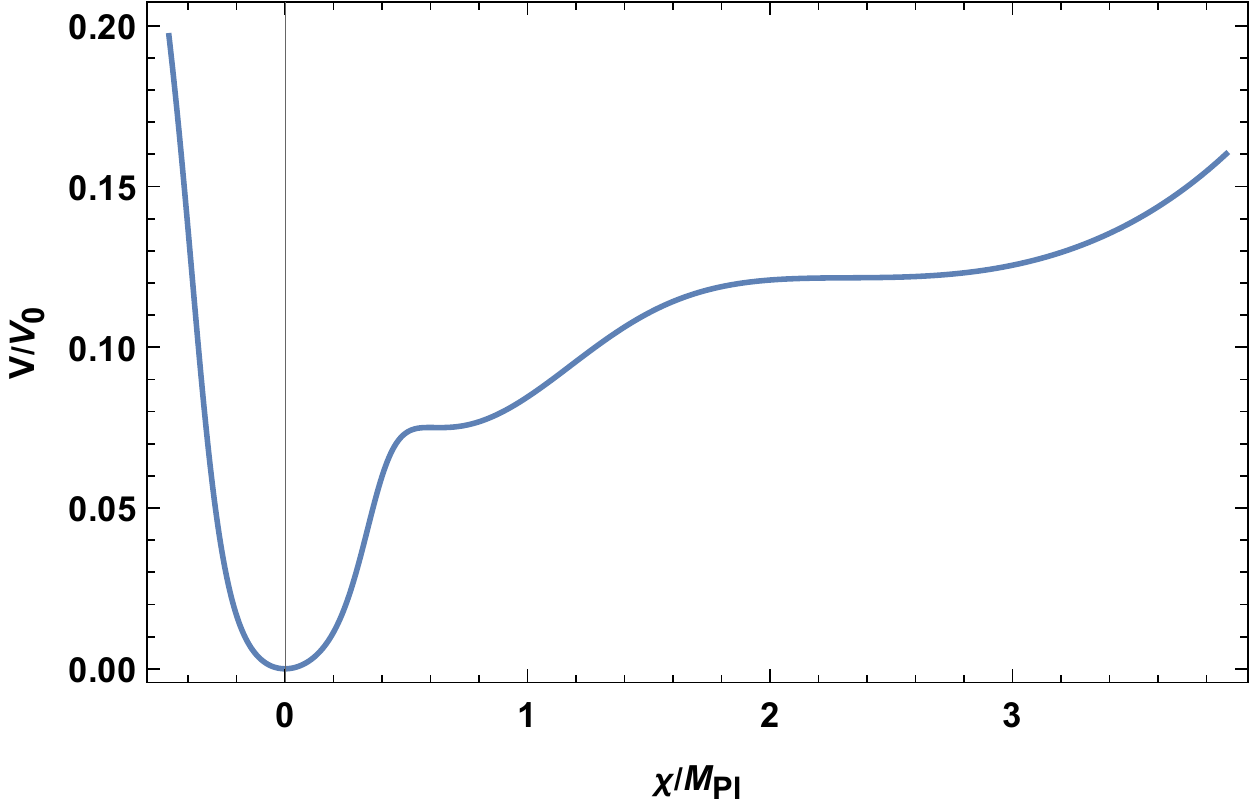}}
	\subfigure[]{\includegraphics[height=0.25\linewidth]{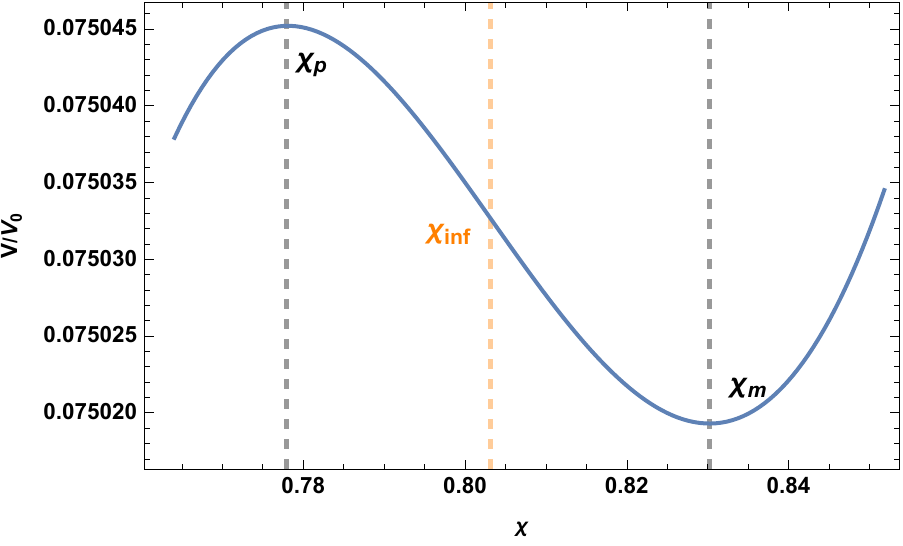}}
	\caption{(a) The potential $V(\chi)$ of the canonical scalar field $\chi$ for the model I with the model parameters as shown in Table \ref{Tab:model}.
	(b) The upward step near the inflection point  $\chi_{\text{inf}}=0.8032 M_{\text{Pl}}$. The power spectrum enhances from $\chi_m=0.8303 M_{\text{Pl}}$ and peaks at $\chi_p=0.7781 M_{\text{Pl}}$. }\label{fig:V-pbh}
\end{figure*}

The inflationary direction is set as $\overline{\varphi}=\varphi$, and  $\overline{T}_i=T_i=c_i/2$. 
In order to study the scalar effective potential, 
we need to define a new canonical scalar field $\chi$ with the transformation ${\rm d}\chi/{\rm d} \varphi=\sqrt{2K_{\varphi\varphi}}$. Therefore, the scalar potential of the inflaton $\varphi$ becomes
\begin{equation}\label{eq:V-pbh}
	V=V_0 \,\frac{\varphi ^2 e^{4 b \varphi ^2} (d \varphi -1)^2}{\left(e^{2 b \varphi ^2}-8 a b \varphi ^2 \left(b \varphi ^2-1\right)\right) \left(2 a \varphi ^2+e^{2 b \varphi ^2} \left(c-\varphi ^2\right)\right)},
\end{equation}
with $c=c_1$, $V_0=M^2/(2 c_2)$ and $d=\lambda/M$. 
By integrating Eq. \eqref{eq:fieldtrans}, 
we can get a generalized relation $\varphi(\chi)$ and the potential $V(\chi)$ numerically as shown in Fig.~\ref{fig:V-pbh}. 
One notes that an inflection point exists in the scalar potential with the addition of the extra exponential term in the K\"ahler potential. 
When the pivotal scale $k_*=0.05~\rm Mpc^{-1}$ leaves the horizon,  
$\chi_*\sim 2.5 M_{\text{Pl}}$ or $\varphi_*\sim0.32 M_{\text{Pl}}$, 
so $e^{-b\varphi^2}$ is small if $b\sim 40$ and the effect of the exponential terms can be neglected.
In this limit, the potential \eqref{eq:V-pbh} reduces to the unmodified potential ~\eqref{eq:v_two}.
In other word, the contribution from the extra exponential term can be ignored in the slow-roll regime.

\begin{figure*}
	\centering
	\subfigure[]{\includegraphics[height=0.25\linewidth]{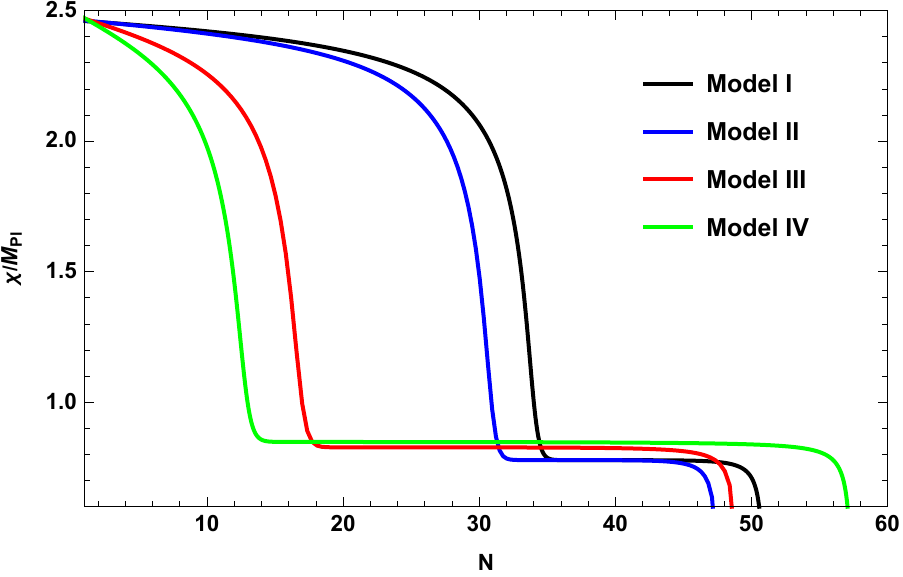}}
	\subfigure[]{\includegraphics[height=0.25\linewidth]{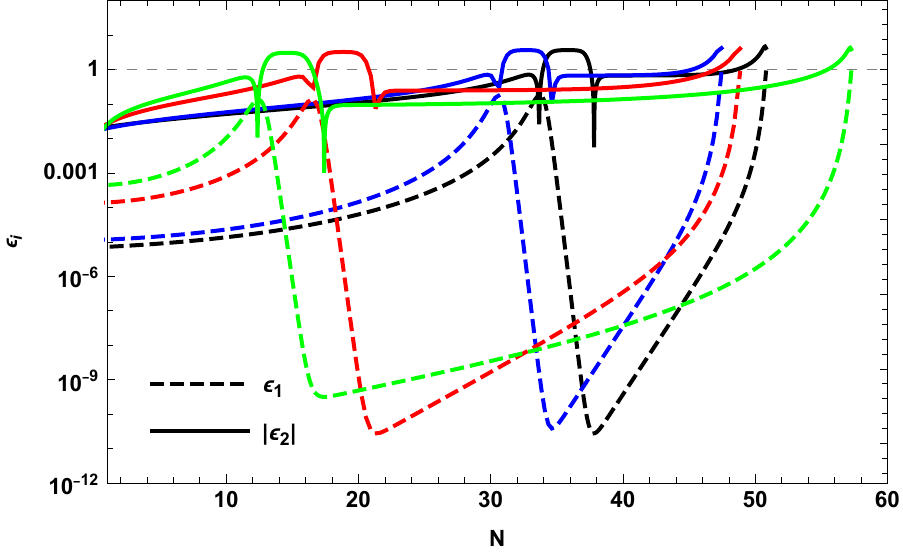}}
	\caption{The evolution of the inflaton $\chi$ (a) and the Hubble slow-roll parameters $\epsilon_1$ and $|\epsilon_2|$ (b) for the models shown in Table \ref{Tab:model}. The inflation ends at $\epsilon_1(\chi)=1$. 
	}\label{fig:chi-slowpara}
\end{figure*}

We show four benchmark points in Table \ref{Tab:model}, 
where the $e$-folding numbers are restricted to be $N\sim 50-60$. 
The parameters $c$ and $d$ affect the total $e$-folding number. 
It is interesting to note that the $e$-folding number spent around the inflection point is about $\Delta N \sim 15-40$, 
this is the reason why we can keep the remaining $e$-folding number before the end of inflation to be $N\sim 50-60$. For the model I, we plot the potential for the inflaton field $\chi$ in Fig.~\ref{fig:V-pbh}. There is an inflection point at $\chi_{\text{inf}}=0.8032 M_{\text{Pl}}$, 
where the slow-roll conditions are no longer satisfied. 
The evolutions of the inflaton $\chi$ and the slow-roll parameters $\epsilon_1$ and $\epsilon_2$ in terms of the number of $e$-folds $N$ are shown in Fig.~\ref{fig:chi-slowpara}. 
The inflation ends at $\epsilon_1 =1$ where the value of inflaton is $\chi_e=0.3936 M_{\text{Pl}}$ for the model I. 
Near the inflection point at $\chi_{\text{inf}}=0.8032 M_{\text{Pl}}$, 
there is a plateau in the evolution of inflaton as a function of the number of $e$-folds $N$,
the slow-roll parameters are $\epsilon_1\sim 10^{-7}$ and $\epsilon_2\sim 3$, and 
the inflation lasts almost 15 $e$-folds. 
This is important for the enhancement of the scalar power spectrum, since the power spectrum under the slow-roll approximation is given by
\begin{equation}
\mathcal{P_{\zeta}}(k)\simeq\frac{1}{12\pi^2}\frac{V(\chi)}{\epsilon_1}.\label{Eq:approxi}
\end{equation}
By varying the parameter $b$, we can get an enhanced $\mathcal{P_{\zeta}}$ and then proper PBH abundance.  However, this approximation in Eq. \eqref{Eq:approxi} fails to give us the accurate power spectrum near the inflection point. Thus, it is necessary to solve the Mukahanov-Sasaki equation numerically to obtain the accurate power spectrum.

\begin{table*}[htp]
	\centering
	\begin{tabular}{m{0.08\textwidth}<{\centering}m{0.09\textwidth}<{\centering}m{0.09\textwidth}<{\centering}m{0.09\textwidth}<{\centering}m{0.09\textwidth}<{\centering}m{0.09\textwidth}<{\centering}m{0.09\textwidth}<{\centering}m{0.09\textwidth}<{\centering}m{0.1\textwidth}<{\centering}m{0.08\textwidth}<{\centering}}
	\hline\hline
	Model& $a$&$b$&$d/A$& $\chi_*(M_{\text{Pl}})$& $\chi_p(M_{\text{Pl}})$ & $\chi_e(M_{\text{Pl}})$ & $n_s$ & $r$ & $N$    \\
	\hline
	 I & -0.480&  46.79179&1.0032&2.463&0.7781& 0.3936&0.9603&$1.1\times10^{-4}$&50.9\\
	II&  -0.480 & 46.7649 &1.0031 &2.466& 0.7782 &0.3935& 0.9652 &$1.8\times10^{-4}$ & 47.5\\ 
	 III&-0.470&43.62443&1.0020&2.485 &0.8272 &0.4043&0.9627&$2.1\times10^{-3}$&48.9 \\ 
	IV& -0.467 & 42.3915 & 1.0 &2.50&0.8481&0.4087  &0.9772 &$6.9\times10^{-3}$ & 57.6\\ 
	\hline\hline
\end{tabular}
	\caption{The model parameters with $c=0.14$  and the predicted observables $n_s$ and $r$.
	$\chi_*$ and $\chi_e$ correspond to the value of the inflaton at the horizon exit and
the end of inflation. $\chi_p$ corresponds to value of the inflaton at which
the power spectrum reaches the peak value.}\label{Tab:model}
\end{table*}

The numerical results for $n_s$ and $r$ at the pivotal scale $k_*=0.05~\rm Mpc^{-1}$ are shown in Table \ref{Tab:model}, 
and they are consistent with the CMB constraints $n_s=0.9649\pm 0.0042~(68\%\rm~ CL)$ and $r_{0.05}<0.036~(95\%\rm~CL)$  \cite{Akrami:2018odb,BICEP:2021xfz}.
The mass and abundance of PBHs are sensitive to the parameters $a$ and $b$. 
For a fixed $c$, a smaller $|a|$ will give a more broad peak for $\mathcal{P_{\zeta}}$ and then produce PBHs with a larger peak mass. 
When $d$ is slightly bigger than $A=\sqrt{27/32c}$, 
the $e$-folding number increases rapidly and $r$ decreases to $10^{-4}$.   
The parameter $V_0$ is fixed by the amplitude of the power spectrum at the horizon crossing.
The constraint on the amplitude of the power spectrum from Planck data is $\mathcal{P_{\zeta}}(k_*)=2.10\times10^{-9}$ \cite{Akrami:2018odb,Aghanim:2018eyx}.

From Fig.~\ref{fig:chi-slowpara},
we see that the number of $e$-folds near the inflection is about $\Delta N \simeq 15$ for the model I and II,
whereas the number of $e$-folds is about $\Delta N\simeq 31$ for the model III and $\Delta N\simeq 40$ for the model IV.
When the inflaton climbs an upward step in the potential from its minimum $\chi_m$ to the maximum $\chi_p$,
the slow-roll parameter $\epsilon_1$ decreases and the power spectrum is enhanced.
The climbing step is shown in the right panel of Fig.~\ref{fig:V-pbh}.
Near the inflection point, the inflaton first decelerates and then accelerates when it rolls over the maximum point $\chi_p$ of the potential that gives the peak of the power spectrum.
After the upward climbing, inflaton goes into the acceleration phase which decreases the power spectrum by the ratio $1/\epsilon_1$ \cite{Inomata:2021tpx}.
As seen from Fig. \ref{fig:chi-slowpara}, 
if $\epsilon_1$ increases more slowly after $\chi_p$,
then the inflaton stays near the inflection longer,
and we get broader power spectrum.
On the other hand, as shown in the right panel of Fig. \ref{fig:chi-slowpara},
the minimum value of $\epsilon_1$ in the model IV is bigger, so its peak value of the power spectrum is expected to be smaller.
\subsection{Primordial curvature perturbations}

Near the inflection point, the inflaton experiences a period of ultraslow-roll inflation and the amplitude of the primordial scalar power spectrum is enhanced.
In order to obtain the power spectrum 
\begin{equation}
\mathcal{P_{\zeta}}=\frac{k^3}{2\pi^2}\left|\frac{u_k}{z}\right|^2
\end{equation}
for the primordial curvature perturbation $\mathcal{R}$,
we numerically solve the Mukhanov-Sasaki equation \cite{Sasaki:1986hm,Mukhanov:1988jd} 
\begin{equation}
\label{Eq:MS}
u_k''+\left(k^2-\frac{z''}{z}\right)u_k=0,
\end{equation}
for the scalar mode $u_k=-z\mathcal{R}$, where the prime means derivative with respect to the conformal time $\tau$, 
$z=\chi'/H$ and
$z''/z$ can be expressed in terms of the Hubble slow-roll parameters as
\begin{equation}
	\frac{z''}{z}= \mathcal{H}^2\left[2-\epsilon_1+\frac{3}{2}\epsilon_2-\frac{1}{2}\epsilon_1\epsilon_2+\frac{1}{4}\epsilon_2^2+\frac{1}{2}\epsilon_2\epsilon_3\right],
\end{equation}
with $\mathcal{H}=aH$. In the limit $k\to\infty$, the mode function $u_k$ takes the solution
\begin{equation}
u_k\to \frac{1}{\sqrt{2k}} e^{-ik\tau}.
\end{equation}

\begin{figure*}
	\centering
	\subfigure[]{\includegraphics[height=0.3\linewidth]{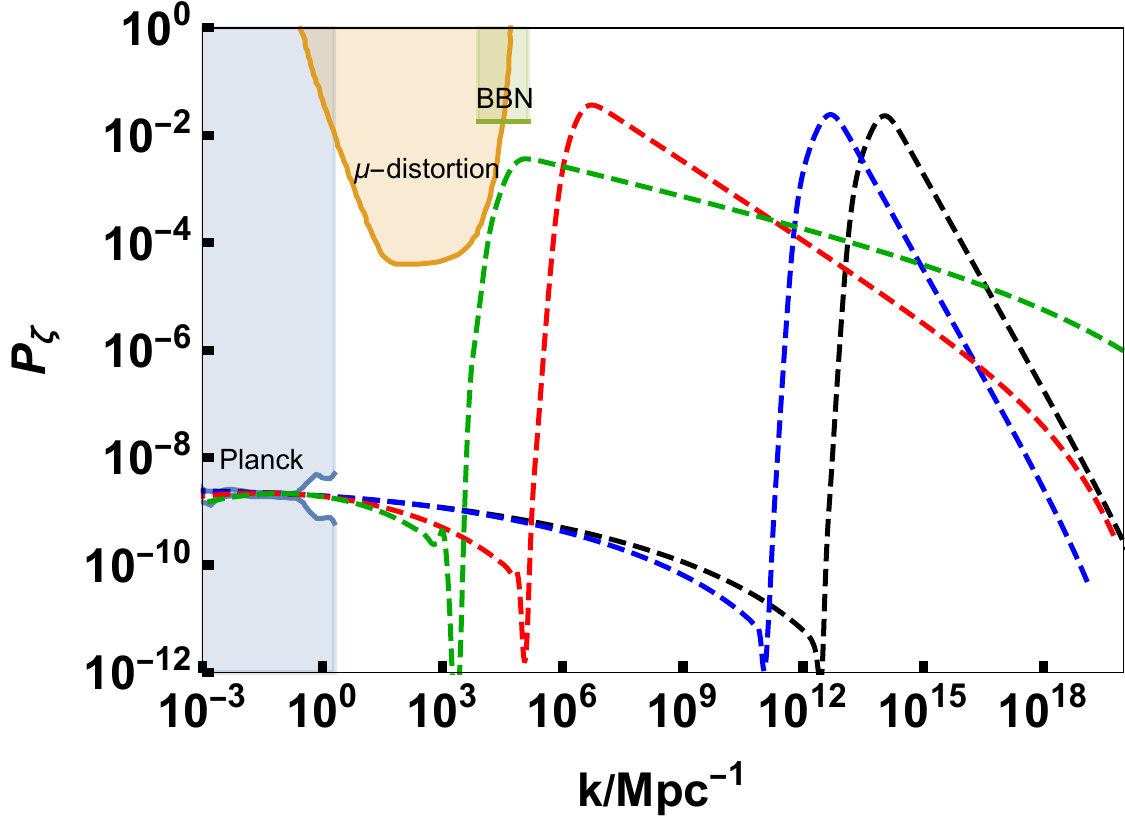}}
	\subfigure[]{\includegraphics[height=0.3\linewidth]{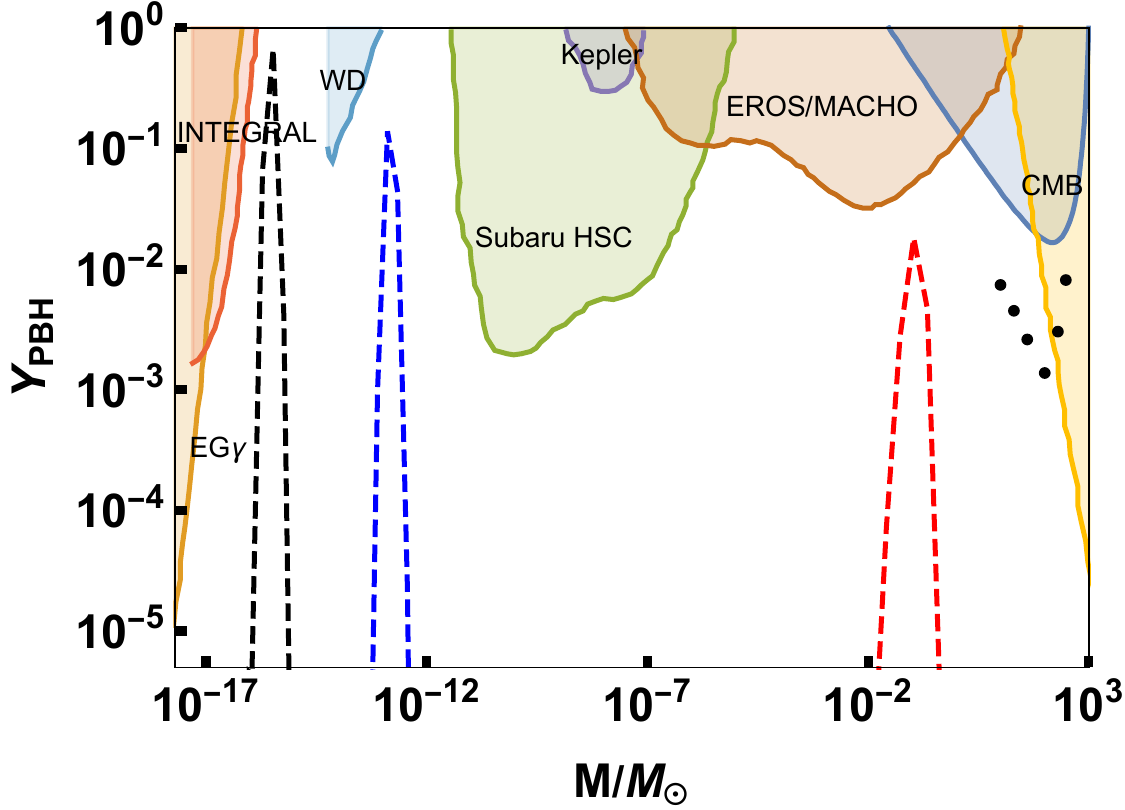}}
	\caption{The scalar power spectrum and PBH abundances. 
	The black, blue, red, and green dashed lines correspond to the models I, II, III, IV, respectively. (a) The shaded constraints are from CMB observations \cite{Akrami:2018odb}, $\mu$-distortion of CMB \cite{Fixsen:1996nj}, and the effect on the ratio between neutron and proton during the big bang nucleosynthesis (BBN) \cite{Inomata:2016rbd}, respectively. (b) The shaded constraints are from the extragalactic gamma rays from PBH evaporation (EG$\gamma$) \cite{Carr:2009jm}, galactic center 511 keV gamma-ray line (INTEGRAL), white dwarf explosion (WD) \cite{Graham:2015apa}, microlensing events with Subaru HSC \cite{Niikura:2017zjd,Cai:2019amo}, the Kepler satellite \cite{Griest:2013esa}, EROS/MACHO \cite{Tisserand:2006zx}, 
	stochastic gravitational wave background by LIGO \cite{Raidal:2017mfl}, LIGO merger rate \cite{Ali-Haimoud:2017rtz}, and accretion constraints from CMB \cite{Ali-Haimoud:2016mbv,Poulin:2017bwe}. The PBH abundance of Model IV is not shown in the figure since the peak abundance is only $10^{-83}$. It was argued that the constraint from WD explosion is not robust and the constraint can be removed \cite{Montero-Camacho:2019jte}.}
	\label{fig:ps_fpbh}
\end{figure*}

We plot the numerical results for the power spectrum in Fig. \ref{fig:ps_fpbh} and show the peak values of the power spectrum in Table \ref{Tab:pbh}. 
The black, red, blue, and green dashed lines correspond to the model I, II, III, and IV, respectively. 
Near the peak point $\chi_{p}$, 
the power spectrum is enhanced to $\mathcal{O}(0.01)$. 
The power spectrum in the models I and II has a sharp peak whereas the power spectrum in the models III and IV has a broad peak. 
The peak point $\chi_p$ in the model III and model IV are further away from the endpoint $\chi_e$ of inflaton than the other two models
and the plateau in the potential of the inflaton $\chi$ in the models III and IV are more broader than the other two models.

\begin{table*}[htp]
	\centering
	\begin{tabular}{m{0.1\textwidth}<{\centering}m{0.15\textwidth}<{\centering}m{0.15\textwidth}<{\centering}m{0.15\textwidth}<{\centering}m{0.15\textwidth}<{\centering}m{0.15\textwidth}<{\centering}}
		\hline\hline
		Model & $k_\text{peak}({\rm Mpc^{-1}})$ & $\mathcal{P_{\zeta}}$ & $M_\text{PBH}( M_{\odot})$ & $Y_\text{PBH}$  & $f_\text{GW}(\rm Hz)$       \\
		\hline
		I& $1.05\times 10^{14}$ & 0.023& $ 3.31\times10^{-16}$ &$0.67$ & $2.09$  \\ 
		II& $4.40\times 10^{12}$ & 0.024 & $ 1.29\times10^{-13}$ &$0.14$  & $ 5.58\times10^{-2}$  \\ 
		III& $5.85\times 10^{6}$ & 0.036 & $ 0.11$&0.018 & $7.08 \times 10^{-8}$  \\ 
		IV&$1.56\times10^{5}$&0.004& 152&$10^{-83}$&$1.51\times10^{-9}$\\
		\hline\hline
	\end{tabular}
	\caption{The peak values of the power spectrum, the mass and abundance of PBHs, and the frequency of SIGWs.}\label{Tab:pbh}
\end{table*}

\subsection{PBH formation}

The enhanced primordial curvature perturbations cause gravitational collapse in the overdense region at the horizon reentry during the radiation dominated era \cite{Carr:2016drx,Fu:2019ttf,Carr:2020gox}. 
If the density fluctuation is larger than a centain threshold $\delta_c$, the gravity can overcome the pressure and hence PBH forms. 
The critical threshold $\delta_c$ for PBH formation has a wide range, 
from 0.07 to 0.7 \cite{Shibata:1999zs,Polnarev:2006aa,Musco:2008hv,Musco:2012au,Harada:2013epa,Germani:2018jgr}. 
In this paper, we take $\delta_c=0.4$  \cite{Tada:2019amh,Escriva:2019phb,Yoo:2020lmg,Solbi:2021wbo}. 
Assuming the primordial perturbations obey Gaussian statistics, 
the fractional energy density of PBHs at their formation time is given by the Press-Schechter formalism \cite{Press:1973iz}
\begin{equation}
\beta(M)\equiv\frac{\rho_\text{PBH}}{\rho_\text{tot}}  \simeq\sqrt{\frac{2}{\pi}}\frac{\sqrt{P_{\zeta}}}{\mu_c} \exp\left(-\frac{\mu_c^2}{2P_{\zeta}}\right),
\end{equation}
where $\mu_c=9\delta_c/2\sqrt{2}$.
The fractional energy density of PBHs  with the mass $M$ to DM is \cite{Carr:2016drx,Carr:2020gox}
\begin{equation}
\begin{split}
Y_\text{PBH}(M)
=&\frac{\beta(M)}{3.94\times10^{-9}}\left(\frac{\gamma}{0.2}\right)^{1/2}\left(\frac{g_*}{3.36}\right)^{-1/4}\\
&\times\left(\frac{0.12}{\Omega_{DM}h^2}\right)\left(\frac{M}{M_{\odot}}\right)^{-1/2},
\end{split}
\end{equation}
where $M_{\odot}=1.99\times10^{30}$ kg is the solar mass, $\gamma=0.2$ \cite{Carr:1975qj}, $\Omega_\text{DM}h^2=0.12$ \cite{Aghanim:2018eyx}. The effective degrees of freedom $g_*$ at the time of PBH formation is $g_*=106.75$ in the radiation dominated era.
The mass of PBHs is 
\begin{equation}
\frac{M(k)}{M_{\odot}}=3.68\left(\frac{\gamma}{0.2}\right)\left(\frac{g_*}{3.36}\right)^{-1/6}\left(\frac{k}{10^6~\rm{Mpc^{-1}}}\right)^{-2}.
\end{equation}

Using the power spectrum obtained in Fig. \ref{fig:ps_fpbh},
we calculate the mass and abundance of PBHs and display the results in Fig.~\ref{fig:ps_fpbh} and Table \ref{Tab:pbh}. 
A wide mass range of PBH is realized in our model, 
and we show four benchmark points where the PBH masses are around $\mathcal{O}(10^{-16}M_{\odot})$, $\mathcal{O}(10^{-12}M_{\odot})$, $\mathcal{O}(10^{-2}M_{\odot})$ and $\mathcal{O}(10^{2}M_{\odot})$. 
The PBHs with masses around $\mathcal{O}(10^{-16}M_{\odot})$ and $ \mathcal{O}(10^{-12}M_{\odot})$ can make up almost all DM and the peak abundances are $Y_\text{PBH}\simeq 1$. 
The PBH with the mass around $\mathcal{O}(10^{-2}M_{\odot})$ only explains part of dark matter with $Y_\text{PBH}\simeq10^{-2}$.
However, from Table \ref{Tab:pbh}, 
we see that the PBH with the mass around $\mathcal{O}(10^2M_{\odot})$ is hard to explain DM because of the significantly small value of $Y_\text{PBH}$.

\subsection{Scalar induced gravitational wave}

Since the scalar perturbations and tensor perturbations are coupled at the second order, 
the large primordial curvature perturbation at small scales will induce second-order tensor perturbations. 
On CMB scale, the amplitude of tensor perturbation is much smaller than that of the scalar perturbation. 
The tensor-to-scalar ratio is constrained as $r_{0.05}<0.036$ (95\% CL) \cite{BICEP:2021xfz}. 
Therefore, the second-order tensor perturbations induced by the enhanced scalar perturbations may be larger than primordial GWs. 
The perturbed metric in the Newtonian gauge is given by
\begin{equation}
\begin{split}
    {\rm d}{s^2}=a^2\left[-(1+2\Phi){\rm d}{\tau^2}
+((1-2\Psi)\delta_{ij}+\frac{1}{2}h_{ij}){\rm d}{x^i}{\rm d}{x^j}\right],
\end{split}
\end{equation}
In the following calculation, we will neglect the anisotropic stress, so $\Phi=\Psi$.
The equation of motion for the tensor mode, $h_k(\tau)$, sourced by the scalar perturbation $\Phi_k(\tau)$, is
\begin{equation}
\label{eq:tensor}
h''_\textbf{k}+2\mathcal{H}h'_\textbf{k}+k^2h_\textbf{k}=4\mathcal{S}_\textbf{k},
\end{equation}
where the source term is given by
\begin{equation}
\begin{split}
    \mathcal{S}_k=&\int\frac{{\rm d}{^3\tilde{k}}}{(2\pi)^{3/2}}e_{ij}(k)\tilde{k}^i\tilde{k}^j \left[2\Phi_{\tilde{k}}\Phi_{k-\tilde{k}}\right.\\
&\left.+\frac{4}{3(1+w)\mathcal{H}^2}(\Phi'_{\tilde{k}}+\mathcal{H}\Phi_{\tilde{k}})(\Phi'_{k-\tilde{k}}+\mathcal{H}\Phi_{k-\tilde{k}})\right],
\end{split}
\end{equation}
$w=p/\rho$, $e_{ij}(k)$ is the polarization tensor.
In the radiation domination, the Bardeen potential $\Phi_\textbf{k}(\tau)=\Phi(k\tau)\phi_k$ and the transfer function is
\begin{equation}
\Phi(x)=\frac{9}{x^2}\left(\frac{\sin{(x/\sqrt{3})}}{x/\sqrt{3}}-\cos{(x/\sqrt{3})}\right),
\end{equation}
where $x=k\tau$, $1/\sqrt{3}$ is the sound speed of the radiation background. The fluctuation $\phi_k$ is related to the primordial curvature perturbation as
\begin{equation}
\langle \phi_{\bm{k}}\phi_{\tilde{\bm{k}}}\rangle=\delta^3(\bm{k}+\tilde{\bm{k}})\frac{2\pi^2}{k^3}\left(\frac{3+3w}{5+3w}\right)^2\mathcal{P_{\zeta}}.\label{eq:ps}
\end{equation}

The power spectrum of the tensor perturbation is defined as
\begin{equation}
\langle h_{\bm{k}}(\tau)h_{\tilde{\bm{k}}}(\tau)\rangle=\delta^3(\bm{k}+\tilde{\bm{k}})\frac{2\pi^2}{k^3}  \mathcal{P}_h(k,\tau). \label{eq:ph}
\end{equation}
After solving Eq. \eqref{eq:tensor} with Green's function method, 
the power spectrum can be written as \cite{Ananda:2006af,Baumann:2007zm,Kohri:2018awv}
\begin{equation}
\begin{split}
    \mathcal{P}_{h}(k,\tau)=&4\int_0^{\infty}{\rm d} v\int_{|1-v|}^{1+v}{\rm d} u\\ &\times \left(\frac{4v^2-(1+v^2-u^2)^2}{4uv}\right)^2\\
    &\times     I_{RD}^2(u,v,x) \mathcal{P_{\zeta}}(kv)\mathcal{P_{\zeta}}(ku),
\end{split}
\end{equation}
and the fractional energy density of the induced GWs is
\begin{equation}
\Omega_\text{GW}(\tau,k)=\frac{\rho_{GW}(\tau,k)}{\rho_{tot}(\tau)} =\frac{1}{24}\frac{k^2}{\mathcal{H}^2} \overline{\mathcal{P}_h(\tau,k)},
\end{equation}
where the overline denotes the oscillation average,
$u=|\bm{k}-\tilde{\bm k}|/k$ and $v=\tilde{k}/k$. 
During the radiation era, the kernel function is \cite{Espinosa:2018eve,Ananda:2006af,Baumann:2007zm,Kohri:2018awv,Lu:2019sti}
\begin{equation}
    \overline{I_{RD}(u,v,x\to\infty)}=\frac{1}{2}\left(\frac{3(u^2+v^2-3)^2}{4u^3v^2x}\right)^2
    \left(I_s+I_c\right),
\end{equation}
where $I_s=\pi^2\Theta(u+v-\sqrt{3})$ and $I_c=\left(\log\left|\frac{3-(u+v)^2}{3-(u-v)^2}\right|-\frac{4uv}{u^2+v^2-3}\right)^2$.

\begin{figure}[!h]
	\centering
	\includegraphics[height=0.4\columnwidth]{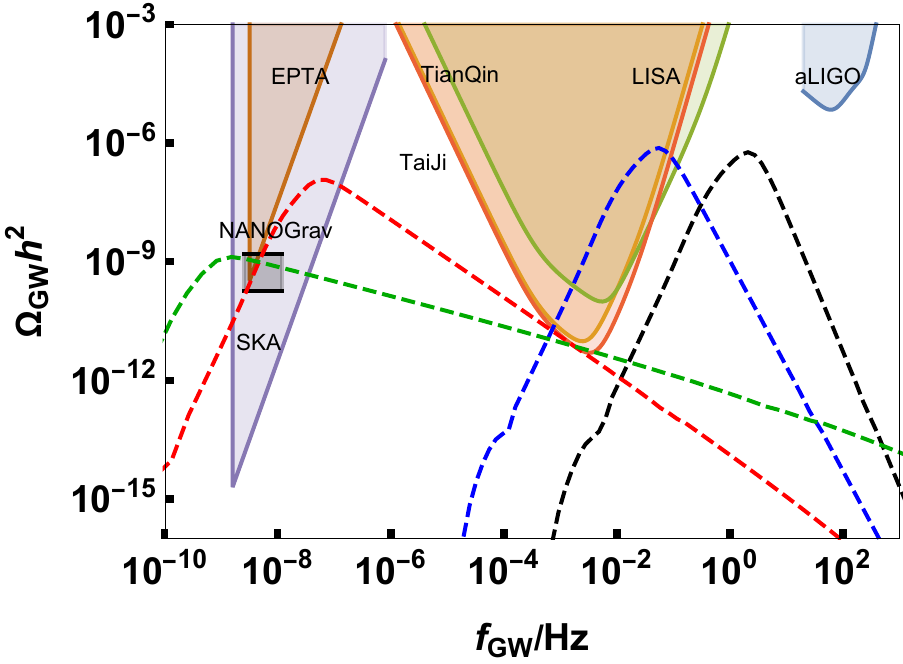}
	\caption{The energy densities of SIGWs. The black, blue, red, and green dashed lines  correspond to the models I, II, III, and IV, respectively. 
	The shaded regions indicate sensitivity of GWs observatories, such as Square Kilometer Array (SKA) \cite{Moore:2014lga}, European PTA (EPTA) \cite{Ferdman:2010xq,Hobbs:2009yy,Hobbs:2013aka,McLaughlin:2013ira}, North American Nanohertz Observatory for GWs (NANOGrav) \cite{Arzoumanian:2020vkk}, Taiji \cite{Hu:2017mde}, TianQin \cite{Luo:2015ght}, Laser Interferometer Space Antenna (LISA) \cite{Danzmann:1997hm}, and Advanced Laser Interferometer Gravitational Wave Observatory (aLIGO) \cite{Harry:2010zz,TheLIGOScientific:2014jea}.}
	\label{fig:gw}
\end{figure}

Converting the wave number to frequency with $f_\text{GW}=9.71586\times10^{-15} k$  Hz/Mpc$^{-1}$, 
the energy density of SIWGs against its frequency is plotted in Fig. \ref{fig:gw}.
From Fig. \ref{fig:gw}, we see that the energy density of SIGWs illustrates some universal dependencies on the frequency \cite{Xu:2019bdp,Cai:2019cdl,Yuan:2019wwo}.
The peak frequencies of these SIWGs are shown in Table \ref{Tab:pbh}.
The SIWGs generated in the models III and IV have a wide peak.
This broad band with frequencies $f_\text{GW}\sim [10^{-10},~10^2]$ Hz, 
comes from the broad enhanced power spectrum with the wavenumber $k\sim[10^{4},~10^{20}]$ Mpc$^{-1}$.
The produced PBH with the stellar mass can constitute part of DM in our universe.
The SIWGs in the nanohertz band can be interpreted as the stochastic GW background
observed by the recent 12.5-yr PTA data released by the NANOGrav \cite{Vaskonen:2020lbd,Kohri:2020qqd}.   
The broad band SIGWs can be tested by multiband GW observations.
The SIWGs generated in the model II have a peak at the frequency $f_\text{GW} \sim \mathcal{O}(10^{-2})$ Hz, which will be tested by the space-based GW detector, such as LISA, Taiji and TianQin.
The SIWGs generated in the model I have a peak around the frequency $f_\text{GW} \sim $ 10 Hz which will be tested by the future ground-based GW observatory.

\section{Conclusion}\label{sec:conclusion}

We have investigated the formation of PBHs and SIGWs from
the general no-scale supergravity inflationary model,
inspired by the string model building.
An inflection point is added by introducing an exponential term into K\"ahler potential.
Thus, the amplitude of the primordial power spectrum is enhanced.
To achieve the enhancement by the inflection point, the model parameters need to be fine-tuned by five decimal digits at most.
The enhanced power spectra of primordial curvature perturbations can have both sharp and broad peaks.
A wide mass range of PBH is realized in our model,
and the frequencies of SIGWs are ranged form nanohertz to kilohertz.
We have shown four benchmark points where the PBH masses are around $\mathcal{O}(10^{-16}M_{\odot})$, $\mathcal{O}(10^{-12}M_{\odot})$ and $\mathcal{O}(10^{-2}M_{\odot})$ and $\mathcal{O}(10^2M_{\odot})$.
The SIGWs accompanied with the formation of stellar mass PBH can interpret the stochastic GW background detected by NANOGrav. 
The PBHs with masses around $\mathcal{O}(10^{-16}M_{\odot})$ and $ \mathcal{O}(10^{-12}M_{\odot})$ can make up almost all DM and the accompanied SIGWs will be testable by the upcoming space-based GW observatories.
The observations of both PBHs and SIGWs can test our model and inflationary physics. 
The broad band SIGWs generated in this model can be tested by multiband GW observations.

\begin{acknowledgments}
This work is supported in part by the National Natural Science Foundation of China under Grants No. 11875062, No. 11875136, and No. 11947302, the Major Program of the National Natural Science Foundation of China under Grant No. 11690021, the Key Research Program of Frontier Science, CAS.
This work was also supported in part by the Natural Science Basic Research Plan in Shanxi Province of China under Grant No. 2020JQ-804,
and  by the Shanxi Provincial Education Department under Grant No. 20JK0685.
\end{acknowledgments}


%

\end{document}